\documentclass{aa}
\usepackage{graphics}
\usepackage{natbib}
\begin{document}
\title{Seyfert galaxies in UZC-Compact Groups}
\author{B.Kelm\inst 1, P. Focardi\inst 1, and V.Zitelli\inst 2} 
\institute{Alma Mater Studiorum - Dipartimento di Astronomia, Via Ranzani 1, 
I-40127 Bologna 
\and
INAF-Osservatorio Astronomico Bologna, Via Ranzani 1 , I-40127 Bologna}
\date{Received 8 September 2003 / Accepted 6 January 2004}
\abstract{
We present results concerning the occurrence of 
Seyfert galaxies in a new automatically selected 
sample of nearby Compact Groups of galaxies (UZC-CGs). Seventeen Seyferts are 
found, constituting $\sim$3\% of the UZC-CG galaxy population. 
CGs hosting and non-hosting a Seyfert member exhibit no significant 
differences, except that a relevant number of Sy2  
is found in unusual CGs, all presenting large velocity dispersion 
($\sigma$$>$400 km\,s$^{-1}$), many neighbours and a high number of 
ellipticals.   
We also find that the fraction of Seyferts in CGs is 3 times as 
large as that among UZC-single-galaxies, and 
results from an excess of Sy2s.
CG-Seyferts are not more likely than other CG 
galaxies to present major interaction patterns, nor to display a bar.      
Our results indirectly support the minor-merging fueling mechanism.   
 }
\titlerunning{Seyferts in UZC-CGs}
\maketitle
%%%%%%%%%%%%%%%%%%%%%%%%%%%%%%%%%%%%%%5
\section{Introduction}
%%%%%%%%%%%%%%%%%%%%%%%%%%%%%%%%%%%%%%%%%%%%
Because of their high number density of galaxies (comparable to the central 
density in clusters) 
and relatively low velocity dispersion ($\approx$ 200-300 km\,s$^{-1}$),  
Compact Groups (CGs) are predicted to constitute the most probable 
sites for strong galaxy-galaxy interactions and mergers to occur.
As a consequence, they are also expected to display 
a high fraction of AGNs, provided they are bound 
systems \citep{Hickson92,Diaferio00} and the interaction-activity 
paradigm \citep{Bar92,shlosman90} holds true. 
The detected fraction of AGNs in CGs might then help to constrain 
the dynamical status of CGs. A high fraction of AGNs would indicate 
that CGs are not only physical, but also highly unstable, and  
would thus support the interaction-activity paradigm,  
as well as hierarchical scenarios in which large isolated elliptical galaxies
\citep{Zablu98,Borne00} are eventually the end-product of every CG.  
Conversely, a low fractions of AGNs would be more in accordance with 
recent results indicating that the occurrence of emission-line galaxies 
decreases in dense environment \citep{Balogh03,Gomez03}  
and with optical and IR observations claiming that spiral galaxies in CGs 
\citep{Sulentic93,Verdes98,kelm03}, in groups \citep{Maia03} and in 
pairs \citep{Bergvall03}  
do not show starburst and/or Seyfert enhancement typically expected in 
interacting galaxies. 
In general, a strong correlation appears to hold between AGN and the 
presence of tidal interaction only for very luminous QSOs \citep{Bahcall97}, 
while the excess of companions for Seyferts is still controversial 
\citep{Dahari85,Keel85,Rafanelli95,Fuentes88,Mackenty89,Keel96,Derobertis98,Schmitt01}.     

Concerning CGs, only the Hickson Compact Groups (HCG, Hickson 1982, 
1997) have been extensively studied as for many years it was the only 
large and uniform sample available. 
The results from this sample are somewhat conflicting.  
Several HCGs show evidence of ongoing interaction, but components 
usually remain distinct, with recognizable morphological types \citep 
{Sulentic97}. 
The fraction of blue ellipticals (which are 
plausible merger remnants) has turned out to be rather low (4 in 55), 
predominantly associated with faint members (Zepf {\it et al.} 1991) and 
similar to the estimated fraction ($\approx$7\%) of currently merging 
galaxies (Zepf 1993). 
Hickson {\it et al.} (1989)
 found the FIR emission in HCGs to be enhanced  compared to a sample 
of field galaxies, but  Sulentic \& de Mello (1993) and Verdes-Montenegro 
{\it et al.} (1998) suggest there is no firm evidence for enhancement.

The specific issue of AGNs in HCGs has been addressed by the Kelm {\it et al.} 
(1998) finding that only $\approx$2\% of the member galaxies display a 
Seyfert spectrum, and that this fraction is similar to 
that found in galaxy pairs. They also find that   
HCG Seyferts are hosted by luminous spirals, as is usually the 
case \citep{Heckman78}. 
However, a relevant population of low-luminosity AGNs 
(LLAGNs) in HCGs as well as in the SCG sample (Iovino 2002)
has been revealed by means of deep resolution spectroscopy 
 (Coziol {\it et al.} 1998a, 1998b, 2000), 
with a significant preference for early-type hosts. 
Coziol {\it et al.} (2000) state that AGNs (including low luminosity/dwarf 
sources) are the most frequent (41\%) activity type encountered in CGs. 
Shimada {\it et al.} (2000) confirm a high fraction of LLAGNs in HCGs, 
but claim that there is no statistically significant difference in 
the frequency of occurrence of emission line galaxies between the HCGs 
and the field.   
Whether a CG environment 
really triggers an AGN at all, or whether 
CGs are favourable hosts for low-excitation, low luminosity 
AGNs only, remains controversial, however. 

In this paper we address the issue of Seyfert occurrence in CGs 
making use of the new large sample of UZC-CGs (Focardi \& Kelm 2002), 
selected from a 3-D magnitude limited  catalogue (UZC, Falco et al. 1999). 
In Sect. 2 the UZC-CG and the Seyfert samples are presented, in \S 3 
UZC-CGs with and without a Seyfert are compared; a similar comparison for 
HCGs is performed in \S 4. In \S 5 the frequency of 
Seyferts in UZC-CGs and in a single-galaxy sample (selected in UZC) 
are discussed. 
In \S 6 and 7 we address the relative occurrence of Sy1, Sy2 and LINERs, 
and in \S 8 the presence of interaction patterns.  

A Hubble constant of $H_{o}$\,=\,100\,km\,s$^{-1}$\,Mpc$^{-1}$ 
is used throughout.
\section {The samples}
UZC-CGs (Focardi\,\&\,Kelm 2002) have been extracted from the UZC 
catalog (Falco et al. 1999) using an objective algorithm.  
UZC lists redshift for nearly 20\,000 galaxies in the northern sky 
and is 96\% complete for m${_B}$\,$\le$\,15.5 galaxies.    
UZC-CGs are systems of 3 or more galaxies  
lying inside a 200$h^{-1}$\,kpc radius area and radial velocity within 
1000\,km\,s$^{-1}$ from the center.  
Possible ACO clusters substructures  have been 
excluded from the UZC-CG sample. 
  
The present analysis is restricted to 192 UZC-CGs (639 galaxies) 
in the 2500-7500 km\,s$^{-1}$ radial velocity range. 
The lower limit in radial velocity  has been set to avoid possible 
Local Supercluster structures and major contamination 
of distances by the effect of peculiar motions.  
The upper limit avoids including a large population of galaxies 
with uncertain morphological classification. 

Seyfert galaxies are identified by cross correlating UZC-CG galaxies 
with the Veron-Cetty\&Veron (2001, V\&V) AGN catalogue. V\&V
is still the largest all-sky available catalogue of bright nearby Seyfert 
galaxies. 
It is neither complete nor homogeneous, nevertheless it is not 
severely affected by survey biases (i.e. there are no large sky regions in 
which the absence of Seyferts can be attributed to the lack of data) and can 
be used at least for a preliminary analysis.

The NED database has further been inspected to 
include any additional AGN and to check V\&V's 
activity classification. It has also been used to assign morphological 
classification, available for 75$\%$ of the galaxies. 
%%%%%%%%%%%%%%%%%%%%%%%%%%%%%%%%%%%%%%%%%%%
\begin{table*}
\begin{center}
\caption{Seyferts in the UZC-CG Sample. }
\begin{tabular}{||r|l|c|c|l|r|r|r|r|l||}
\hline
 UZC-CG  & name  &V\&V & NED  & morphology & cz  & m$_B$ &lum. rank$^1$ &inter.$^2$ &other id. \cr
\hline
     &               &     &     &        &      &     &   &   & \cr 
 19  & NGC449    & Sy2 & Sy2 & S      & 4750 & 15.2 &3&  & MKN 1 \cr
 23  & NGC513    & Sy2 & Sy2 & Sb/c   & 5840 & 13.4 &1&  & ARK 41\cr 
 30  & UGC1479       & Sy2 & Sy2 & Sc & 4927 & 14.8 &2&  &       \cr
 57  & UGC3752       & Sy2 &     & S? & 4705 & 14.5 &2&  &       \cr
 109 & ZW63.06       & Sy2 &     &    & 3988 & 15.5 &3&  &       \cr
 143 & NGC3798       & Sy1 &     & SB0    & 3563 & 13.9 &1&  &       \cr
 144 & NGC3822   & Sy2 & Sy2 & Sb     & 6138 & 13.7 &1& Y & HCG58a \cr  
 156 & NGC4074       & Sy2 & Sy2 & S0 pec.& 6802 & 15.4 &6&  & AKN347 \cr
 162 & NGC4169       &     & Sy2  & S0/a   & 3755 & 12.9 &1&    & HCG61a\cr
 186 & IC4218        & Sy1 & Sy1 & Sa     & 5808 & 14.4 &1&  & \cr
 200 & ZW45.099      & Sy2 & Sy2 &        & 6870 & 15.4 &2& & \cr 
 207 & NGC5395   &     & Sy2 & Sb     & 3522 & 12.6 &4& Y  & Arp84, IZw77 \cr
 240 & NGC5985       & Sy  & Sy1 & Sb   & 2547 & 12.0 &1&  & \cr 
 242 & NGC5990   & Sy2 &     & Sa pec.& 3726 & 13.1 &1& &  \cr
 272 & NGC6967       & Sy2 & Sy2 & SB0    & 3768 & 14.3 &3&  & \cr
 276 & UGC11950  & Sy2 &     & E      & 6224 & 14.3 &1&   & \cr
 279 & NGC7319   & Sy2 & Sy2 & SBpec. & 6652 & 14.8 &2& Y   &HCG92c \cr
     &               &     &     &        &      &     &   &   & \cr 
\hline
\end{tabular}
\note{col.8 indicates the luminosity rank of the Sy galaxy within its group} 
\note{col.9 indicates the presence of major morphological-peculiarities/interaction-patterns visible on POSS plates.}  
\par
\end{center}
\end{table*}            
%%%%%%%%%%%%%%%%%%%%%%%%%%%%%%%%%%%%%%%%%%%%%%%
In Table 1 basic data for each Sy1 and Sy2 in UZC-CGs are listed:  
UZC-CG number (column 1), Seyfert name (column 2), V\&V
and NED activity classification (column 3 and 4), 
morphological classification (column 5),
radial velocity (column 6), apparent blue magnitude as in UZC (column 7), 
luminosity rank
(1 = first ranked, 2 = second ranked ...) (column 8), 
presence of interaction pattern (column 9), other identification (column 10). 

To identify LINERs (L) we have also inspected the list by 
Carrillo et al. (1999). LINERs are named Sy3 in V\&V, and occasionally 
AGN in NED. LINERs in UZC-CGs are listed in Table 2. 
LINERs are found in galaxies of earlier Hubble type than Seyferts 
and their nuclear continua are usually dominated by old stars 
\citep {Heckman80,Ho03}. We have kept Sy1 and Sy2 separate from LINERs 
because the nature of the central power-source in LINERs remains uncertain; 
they might constitute a transition class between 
non thermal objects and starburst \citep{Carrillo99} or between 
normal galaxies and Seyferts rather than a true low luminosity extension 
of the AGN sequence. 
%%%%%%%%%%%%%%%%%%%%%%%%%%%%%%%%%%%%%%%%%%%%%%%%%%%%%%%%%%%%%%%%%
\section{CGs with and without a Seyfert: is there any difference?}
%%%%%%%%%%%%%%%%%%%%%%%%%%%%%%%%%%%%%%%%%%%%%%%%%%%%%%%%%%%%%%%%%%%
\begin{figure}
\resizebox{\hsize}{!}{\includegraphics{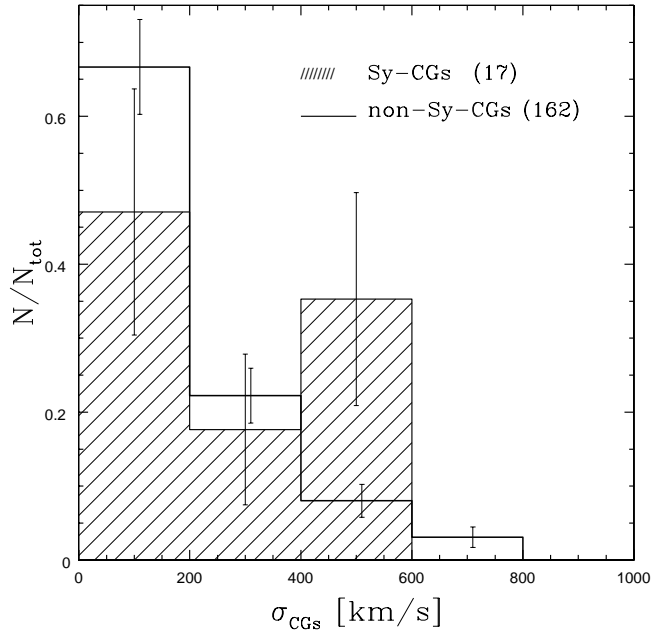}}
\hfill
\caption[]{
Velocity dispersion distribution 
for Sy-CGs (hatched) and non-Sy-CGs.
Though an excess of Sy-CGs emerges in the 400-600 km\,s$^{-1}$ bin, 
differences between distributions are not highly significant 
(91\% c.l. according to the KS test). 
}
\end{figure} 
%%%%%%%%%%%%%%%%%%%%%%%%%%%%%%%%%%%%%%
\begin{table*}
\begin{center}
\caption{LINERs in the UGC-CG Sample. }
\begin{tabular}{||r|l|c|c|l|r|r|r|r|l||}
\hline
 UZC-CG   & name  &V\&V & NED/Carrillo  & morphology & cz & m$_B$ &lum. rank & inter &other id. \cr
\hline
     &               &     &     &        &      &     &   &   & \cr 
  4  & NGC51         &     & L   & S0 pec.& 5342 & 14.6 &1&   & \cr 
 14  & NGC315    & Sy3   & L   & E      & 4956 & 12.5 &1&   & \cr 
 16  & NGC410 &     & L   & E      & 5294 & 12.6 & 1&  & \cr
 25  & NGC536 &     & AGN & SB     & 5251 & 13.2 &2&   &HCG10a \cr 
 100 & NGC2911 & Sy3 & Sy  & S       & 3217 & 13.6 &1& Y &ARP232 \cr    
 162 & NGC4175 &     & AGN & Sbc    & 4019 & 14.2 &2& Y & HCG61c,KTG42 \cr 
 169 & NGC4410 & Sy3 & L & Sab     & 7601 & 13.6 & 1& Y &MKN1325 \cr   
 203 & NGC5318 &     & AGN & S0?    & 4329 & 13.5 & 1 & Y &   \cr
 234 & NGC5851 & Sy3 & L   & S?     & 6470 & 14.9 &2& Y   &   \cr 
 240 & NGC5982 &   & L   & E      & 2918 & 12.4 & 2 &   &   \cr
 252 & NGC6286 &     & L & Sbpec. & 5551 & 14.2 &1 &Y&Arp 293 \cr
     & NGC6285       &     & L & S0     & 5691 & 14.6 & 2&    & \cr
 257 & NGC6482       &     & L   & E      & 3913 & 12.8 &1&     & \cr
 259 & NGC6500 & Sy3 & L   & Sab    & 2986 & 13.4 & 1 &     & \cr
 284 & NGC7549       &     & AGN & SBcd   & 4716 & 14.1 & 2&Y&HCG93b,Arp99 \cr
     & NGC7550 &     & AGN  & S0   & 5072 & 13.9 & 1&     &HCG93a \cr
 290 & NGC7769 &     & L    & Sb   & 4197 & 12.9 &1&    &   \cr 
     &               &     &     &        &      &     &   &   & \cr 
\hline
\end{tabular}
\par
\end{center}
\end{table*}
%%%%%%%%%%%%%%%%%%%%%%%%%%%%%%%%%%%%%5
To derive useful constraints on the role of the environment on 
AGN activation we investigate whether a segregation can be 
found between CGs hosting (Sy-CGs) or not (non-Sy-CGs) 
a Seyfert member. 
We find 17 Sy-CGs and 162 non-Sy-CGs. The 13 CGs hosting a LINER/Sy3/AGN 
have been excluded from our UZC-CG sub-sample.   

Figure 1 shows the velocity dispersion distributions 
of CGs with (hatched) and without (solid line) a Seyfert member. 
Distributions peak below 200 km\,s$^{-1}$ in both samples  
however, Sy-CGs  
are marginally more likely (at 91\% confidence level according to the 
KS test) to display larger velocity dispersion. 
This could indicate that CGs hosting a Seyfert  
are systems more massive than those without. 

Figure 2 shows the distribution of the number of large scale neighbours  
(number of galaxies within  1$h^{-1}$Mpc radius and 
$|$$\Delta$$cz$$|$$\leq$1000\,km\,s$^{-1}$ from the CG center) 
for both samples. 
%%%%%%%%%%%%%%%%%%%%%%%%%%%%%%%%%%%%%%%%%%%5
\begin{figure}
\resizebox{\hsize}{!}{\includegraphics{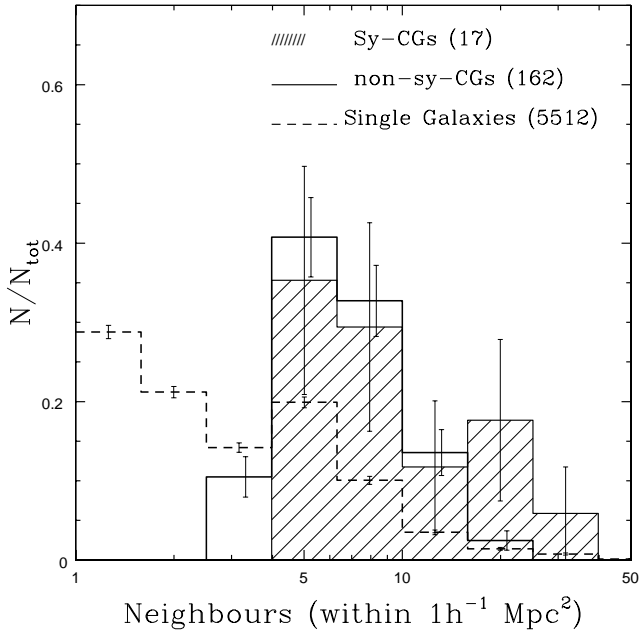}}
\hfill
\caption[]{
Neighbour (within 1\,h$^{-1}$\,Mpc) distributions
for Sy-CGs (hatched) and non-Sy-CGs. 
The difference between the distributions is significant
(98\% c.l. according to the KS) indicating that Sy-CGs are more likely to 
be embedded in dense large scale environments. 
The neighbour distribution of UZC-galaxies presenting no close companion/s 
on the CG scale (single-galaxies) is also plotted.     
}
\end{figure} 
%%%%%%%%%%%%%%%%%%%%%%%%%%%%%%%%%%%%%%%%%%%%%%
The KS-test indicates that Sy-CGs are more likely 
(at 98\% c.l.) than non-Sy-CGs to be associated with a  
large number of companions i.e. they reside in a 
denser large-scale environment. 
Also shown in Fig.2 is the large-scale neighbour distribution 
of UZC-single-galaxies (i.e. galaxies with no neighbour within 
200\,$h^{-1}$\,kpc, see also \S 5), indicating that CGs 
(either with or without a Seyfert member) are embedded in systems 
denser than the environment of galaxies presenting no close 
neighbour.  

Figure 3 shows the morphological distribution of galaxies in Sy-CGs 
(squares) and non-Sy-CGs (triangles).
The differences are not significant, 
yet Fig. 3 suggests that Sy-CGs might include a population
rich in ellipticals and deficient in late spirals. 
Among Seyferts themselves, only one is an elliptical and  
accordingly the data indicate that Seyferts tend to be common when the 
fraction of late spirals among companions is low.
The result is fully consistent with a scenario predicting that Seyferts 
are more common in large and rich groups \citep {Monaco94}. 

If galaxies rather than CGs were to be compared,   
the differences in Fig. 1 and Fig. 2 would become significant at the  
3$\sigma$ level. 
%%%%%%%%%%%%%%%%%%%%%%%%%%%%%%%%%%%%%%%%%%%%%%%%%     
\begin{figure}
\resizebox{\hsize}{!}{\includegraphics{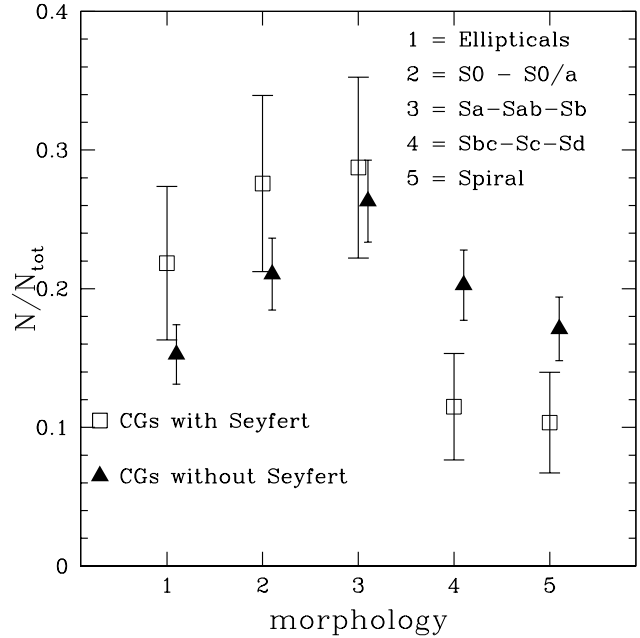}}
\hfill
\caption[]{
Morphological distributions of galaxies in Sy-CGs (squares) and non-Sy-CGs 
(triangles). 
Distributions are not significantly different, however a trend possibly emerges
indicating that late spirals are lacking in Sy-CGs. }
\end{figure} 
%%%%%%%%%%%%%%%%%%%%%%%%%%%%%%%%%%%%%%%%%
Nevertheless, all differences between the samples are induced 
by having included in the Sy-CG sample a few early-type rich CGs 
which display a large velocity dispersion, a large number of neighbours 
and an excess of ellipticals. When the 20 CGs with the lower and the 20 
CGs with the higher velocity dispersion are excluded from the UZC-CG sample 
the differences in $\sigma$ and 
neighbour-density between both populations decrease below the 60\% c.l., 
and at the same time, halves the number of ellipticals in the Sy-CG sample.   
Hence, the present analysis indicates  
that Sy-CGs and non-Sy-CGs are not significantly  
different, except that 5 Sy2 are 
found in CGs which are unusual in terms of $\sigma$, neighbour and 
elliptical content, and closely resemble loose groups. 
%%%%%%%%%%%%%%%%%%%%%%%%%%%%%%%%%%%%%%%%%%%%%%%%%%
\section {Seyferts in HCGs}
%%%%%%%%%%%%%%%%%%%%%%%%%%%%%%%%%%%%%%%%%%%%%%%%%%%%%%%%5
To check whether any differences between systems hosting or not  
a Seyfert member might be found in the HCG sample,  
we next compare the velocity dispersion and 
morphological content of HCGs hosting and not hosting a Seyfert.  
To be consistent with the selection criteria adopted for UZC-CGs, 
only the 31 HCGs in the radial velocity range 2500-7500 km\,s$^{-1}$ are 
considered here.  
Seyferts in HCGs are identified according to the activity classification 
available in V\&V. 7 HCGs are found to host a Seyfert:  
the total fraction of Seyferts in this HCG subsample is 
$\sim$6\%.

Figures 4 and 5 show that Sy-HCGs and non-Sy-HCGs display similar 
velocity dispersions as well as similar morphological compositions, 
although there might be a tendency towards Seyferts preferring low-$\sigma$, 
spiral-rich HCG hosts.  
This confirms that, in general, the presence of a Seyfert is not linked to 
specific CG properties. We stress that the HCG sample lacks the extreme 
CG population (high velocity dispersion, many neighbours and a significant  
early-type galaxy content) present in UZC-CGs, which we have shown to 
be responsible for the difference between Sy-CGs and non-Sy-CGs samples. 

The fraction of Seyferts in the 2500-7500 km\,s$^{-1}$ HCG sample 
is larger than the fraction we find in UZC-CGs. 
This is certainly due to the extensive spectroscopy performed on HCGs. 
However, HCGs are also biased towards luminous galaxies compared to 
UZC-CGs \citep{kelm03b} because they are selected according to a surface 
brightness enhancement criterion ($\mu$$_G$$<$26) which led Hickson to 
select less compact groups of very luminous galaxies. Nearby HCGs include 
many bright spirals, which are the most likely hosts of Seyferts. 
The fraction of Seyferts in UZC-CGs actually rises to 8\% when restricting 
computations to the brightest spirals only.  

The Seyfert fraction on the whole HCG sample (383 galaxies) 
drops to $\sim$2.6\%, a value close to the one previously reported in Kelm et al. (1998). 
This fraction is much lower than fractions quoted by Coziol et al. (2000) and 
Shimada et al. (2000)  as they have performed accurate deep spectroscopy  
allowing them to detect extremely faint emission lines.  
Even excluding LLAGNs, Coziol et al. (2000) quote (in their Table 6) 
a 18\% AGN fraction in HCGs and a 19\% AGN fraction in SCGs, but these 
fractions still include a large number of LINERs.   
Shimada et al. (2000) claim that a high AGN fraction is not typical of 
HCGs alone, as a similar AGN fraction is found in a comparison field sample.   
The analysis by Ho et al. (1997)  
has revealed that 43\% of the objects in their complete sample 
(486 galaxies with m$_B$$\leq$12.5) could be classified as AGN. 
This indicates that active galaxies are common, and indirectly 
disfavors any AGN fueling mechanism that is seldom observed.
Likewise, it appears rather usual for LLAGNs 
to be associated to early-type hosts \citep {Ho97,Kauffmann03}  
whatever the environment of the galaxy, suggesting that the 
association between AGNs and early-type galaxies in CGs 
\citep{Coziol00,shimada} is not a specific feature of CGs. 
%%%%%%%%%%%%%%%%%%%%%%%%%%%%%%%%%%%%%%%%%%%%%%%%%%%%%%%%%%%5
\begin{figure}
\resizebox{\hsize}{!}{\includegraphics{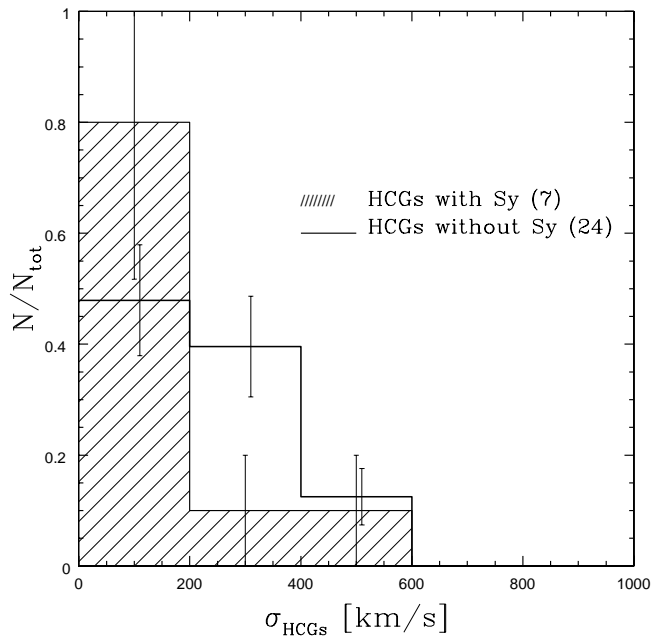}}
\hfill
\caption[]{
Velocity dispersion distributions
for HCGs including (hatched) and non including a Seyfert member.
The difference between the distributions is only marginally significant
(92\% c.l. according to the KS test) indicating that HCGs with a Seyfert
typically display a low velocity dispersion.
}
\end{figure} 
%%%%%%%%%%%%%%%%%%%%%%%%%%%%%%%%%%%%%%%%%%%%%%%%%%%%%%%%%%%%%%%%%555
\begin{table*}
\begin{center}
\caption[] {Seyferts in UZC, UZC-CGs and UZC-single-galaxy samples.}
\begin{tabular}{|l||rrrrc||}
\noalign{\smallskip}
\hline
\noalign{\smallskip}
sample &N$_{tot}$ & N$_{Sy}$(\%) & N$_{Sy1}$ &N$_{Sy2}$ & Sy2:Sy1 \cr 
\hline
\noalign{\smallskip}
UZC-all           & 8488  & 98 (1.2\%) & 33 & 65 & 2:1 \\
UZC-CGs           & 639   & 17 (2.7\%) & 3  & 14 & 5:1 \\ 
UZC-single        & 5512  & 55 (1.0\%) & 22 & 33 & 3:2 \cr 
\noalign{\smallskip}
\hline
\end{tabular}
\end{center}
\end{table*}
%%%%%%%%%%%%%%%%%%%%%%%%%%%%%%%%%%%%%%%%%%%%%%%%%%%%5
\begin{figure}
\resizebox{\hsize}{!}{\includegraphics{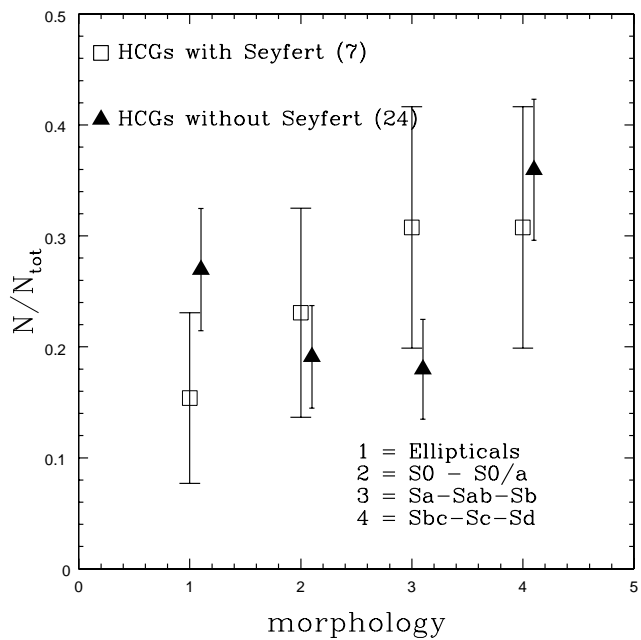}}
\hfill
\caption[]{
Morphological distributions
for HCGs including (squares) and non including (triangles) a Seyfert member. 
Distributions are not significantly different, however a trend 
possibly emerges indicating that in Sy-HCGs ellipticals are underabundant 
while early-spirals being overabundant. 
}
\end{figure} 
%%%%%%%%%%%%%%%%%%%%%%%%%%%%%%%%%%%%%%%%%%%%%%%%%%%%%%%%%%%%%
\section{Are CGs a preferential site for Seyfert galaxies?}
%%%%%%%%%%%%%%%%%%%%%%%%%%%%%%%%%%%%%%%%%%%%%%%%%%%%%%%%%%%%%%%5
In UZC-CGs 17 Seyferts are found, giving a total fraction of
Seyferts over galaxies of 2.7 \%.  For comparison Huchra \& Burg (1992) find 
a 2\% Sy fraction in the CFA1 survey, Hao \& Strauss (2003) a 5\% Sy 
fraction in the SDSS, and Ho et al. (1997) a 11\% Sy fraction in a 
complete (m$_B$$\leq$12.5) RSA subsample. 
Maia et al. (2003) report a 3\% Sy fraction in their SSRS2 sample. 
The Sy fraction in UZC-CGs slightly rises when computation is 
restricted to spiral hosts (3.3\%) or to the brightest galaxy subsample (8\%), 
indicating that Seyferts are commonly associated to bright spiral hosts. 
Counting also LINERs/Sy3 (see Table 2) in UZC-CGs  
enhances the AGN fraction to 5.5\%.

To evaluate whether the fraction of Seyferts in UZC-CGs is 
large, we have cross-correlated the V\&V and the UZC catalogue  
in the 2500-7500 km\,s$^{-1}$ radial velocity range. 
We find 98 galaxies out of 8488 ($\sim$1.2\%) to be  Sy1 or Sy2.  
To further investigate whether the excess of Seyferts in the UZC-CG sample 
is related to their locally high density we have  selected in UZC all galaxies 
that are single on the CG scale (i.e. no close neighbour/s   
within 200$h^{-1}$\,kpc and $|$$\Delta$cz$|$$\leq$1000\,km\,s$^{-1}$). 
In UZC there are 5512 single galaxies; 55 ($\sim$1\%) of these are 
Sy1 or Sy2 in V\&V. Table 3 lists the total number and relative fraction of 
Seyferts in UZC, in UZC-CGs and in the UZC-single-galaxy samples. 
The fraction of Seyferts in UZC-CGs is clearly larger  
than the fraction of Seyferts found in the UZC-single-galaxy 
sample. 

However, one should consider that in the UZC (2500-7500 km\,s$^{-1}$) sample  
most Seyferts (56\%) are single galaxies, while only 17 (20\%) are in CGs, 
meaning that in a generic Seyfert sample, at least half of the galaxies 
display no close neighbour/s. 
Further, because the absence of close companions seems strongly 
related to the number of large-scale neighbours (as shown in Fig. 2), 
many of the single Seyferts are expected to exhibit only few distant 
neighbours. Indeed, nearly half of the single Seyferts (24 in 55) have  
zero/one galaxy out to a 1\,$h^{-1}$Mpc distance. 
The preference for low-density environments displayed by Seyferts    
is fully consistent with the outcome from new large redshift 
surveys such as the 2dFGRS \citep{Lewis02,Balogh03,Bower03} and the SDSS 
\citep{Gomez03,Hogg03,Kauffmann03}, all indicating that emission-line 
galaxies (SB+AGNs) are less frequent in dense environments, and that this 
does not simply correspond to a lack of emission-line galaxies in clusters. 

The number of Seyfert galaxies in UZC-CGs is only 17. Although this 
number is larger than in the HCG subset we have to be careful 
in interpreting observational results based on such small number statistics.  
It is interesting therefore that 
the fractions of Seyferts in UZC-CGs and in 
UZC-single-galaxy systems are rather consistent with the outcome of 
Maia et al. (2003). The authors report a 44\% isolated Seyfert 
fraction and a 28\% triplet+group fraction in the SSRS2, though their 
systems are defined according to criteria different from ours.

A large number of isolated Seyferts is not compatible 
with strong galaxy interactions \citep{Derobertis98,shlosman90} 
being a common AGN-triggering mechanism. 
Conversely, it provides indirect support to 
the minor-merger-AGN-triggering scenario \citep{taniguchi99}.  
%%%%%%%%%%%%%%%%%%%%%%%%%%%%%%%%%%%%%%%%%%%%%%%%%%%%%%%%%%%%%%%55 
\section{Sy1 and Sy2 in CGs and in the single-galaxy sample}
%%%%%%%%%%%%%%%%%%%%%%%%%%%%%%%%%%%%%%%%%%%%%%%%%%%%%%%%%%%%%%%%5
In our UZC-CG sample we find 14 Sy2 and 3 Sy1 galaxies (see Table 3). 
Likewise, among the 7 Seyferts in the HCG subsample (see \S 4) 
only 1 is a Sy1. 
A similar paucity of type 1 Seyferts is found in 
the SCG sample (Coziol et al. 2000).    
In the UZC-CGs the Sy2:Sy1 ratio appears rather high, as values reported in 
the literature  range from 1:1 \citep{Huchra92,Rush93} to 4:1 
\citep{Maiolino95}. Maia et al (2003) find a ratio of 3:1, Hao \& Strauss 
(2003) using data from the SDSS claim that narrow-line AGNs are in a ratio 
of 2:1 to broad-band AGNs. 

We stress that the high ratio of Sy2:Sy1 detected in UZC-CGs is not a generic 
property of the UZC sample. The Sy2:Sy1 ratio in UZC 
(in the radial velocity range 2500-7500 km\,s$^{-1}$) is 2:1,  
comparable to the ratios found in other samples.  
In the single-galaxy sample there are 33 Sy2 and 22 Sy1, 
strongly suggesting that the Sy2:Sy1 ratio is not only a function of 
luminosity \citep{Hao03} but also of environment, 
the two parameters possibly being related \citep{Schmitt01}. 

The excess of Seyferts in CGs discussed 
in the previous section turns out therefore to be an excess of 
Sy2 only. 
In general, a dense environment is not typical of Seyferts; while Sy1s 
avoid dense systems, $\sim$25\% Sy2 are in CGs. A similar conclusion 
was reached by Laurikainen \& Salo (1995) and it is not clear 
whether this is responsible for the excess of companions generally associated 
with Sy2 and not with Sy1 \citep{Dultzin99}  
or whether it might be linked to the idea that two different Sy2 populations 
exist \citep{Storchi01,Krongold02}. 
The asymmetry between Sy1 and Sy2 in CGs does not rule out 
the unified AGN model \citep{Antonucci85,Antonucci93}, 
which claims the Sy1-Sy2 dichotomy to be solely due to the viewing angle. 
It has been suggested that differences in the covering factor of the 
dusty material arise as a consequence of differences in the star-formation 
history of galaxies \citep{Malkan98,Oliva99,Mouri02}. If this is the case, 
the excess of Seyfert 2s in dense galaxy systems simply relates to an excess 
of circumnuclear starbursts in CG galaxies. Spatially resolved optical 
spectra of Seyferts in UZC-CGs could 
confirm whether CG-Sy2 do indeed present a circumnuclear starburst.
\section{LINERs in CGs}
In the UZC-CGs there are 17 galaxies (in 15 CGs) classified as 
LINERs/Sy3/AGN (Table 2). 
LINERs typically have lower nuclear luminosities than 
Seyferts (Heckman 1980, Kauffmann et al. 2003), and it appears that the 
stellar populations in the nuclei of weak AGN are older than in Sy2 
 \citep{Cid01,Joguet01,Ho03}. Schmitt  (2001)  claims  the percentage of 
galaxies with nearby companions to be higher among LINERs than among 
Seyfert galaxies, with differences however disappearing when only 
equal-morphology galaxies are compared. 
 
If we group together CGs including LINERs 
and Seyferts (30 CGs) in the comparison performed in section 3, the 
differences between CGs hosting and not hosting an active galaxy  
become more significant (96\% c.l. when comparing $\sigma$ and 98\% c.l. when 
comparing the neighbour density), suggesting that LINERs tend to be 
found in richer and more massive groups. This corresponds  
to LINERs being more likely associated with early-type galaxies, along 
with the tendency for all luminous galaxies to display some emission 
line when deep spectroscopy is achieved. 

There are many studies in the literature that indicate a positive 
correlation between AGN power and host galaxy luminosity; more powerful 
AGNs are located preferentially in more massive host galaxies. 
The brightest galaxies in UZC-CGs could therefore be expected to be more 
likely to host a Seyfert/LINER. However, when considering the luminosity 
rank (col. 8) of Seyferts it emerges that only 8 are first ranked.  
Sy1 are always first ranked galaxies, but the statistics are clearly 
insufficient to claim the result to be significant.   
LINERs in UZC-CGs appear more likely to be first ranked galaxies 
(11 in 15 CGs) than Sy2 (5 in 11).
It remains then to be clarified if the difference in luminosity rank between 
Sy1, Sy2 and LINERs is simply a consequence of the luminosity/morphology 
segregation or whether and to which level the environment might have 
played a role in causing the segregation. 
%%%%%%%%%%%%%%%%%%%%%%%%%%%%%%%%%%%%%%%%%%%%%%%%%%%%%%%%%%%%
\section{Interaction patterns in CGs}
%%%%%%%%%%%%%%%%%%%%%%%%%%%%%%%%%%%%%%%%%%%%%%%%%%%%%%%%%%%
It has often been speculated that star-formation and AGNs are 
both triggered by close companions.   
Recent results \citep{Lewis02,Gomez03,Hogg03} do however suggest that a 
locally dense environment seems to suppress processes responsible 
for emission-line spectra, and that this holds true also in the field. 
It appears that a companion efficiently triggers SB-AGN phenomena in 
a galaxy only in a low density environment, i.e. in an 
isolated pair \citep{Lambas03,Tanvuia03,Sorrentino03,Bergvall03}.  
This possibly explains why ULIRGs, which are nearly all ongoing 
mergers between spirals, are typical field systems rather than CG members. 
It is still unclear whether a dominant AGN is needed at all to fuel these 
sources or whether they will ever go through an optical QSO phase 
\citep{Genzel01,Tacconi02,Farrah03,Ptak03}.   

Consequently, one might expect two concurrent 
mechanism to operate in CGs, both caused by companions;  
one triggering the AGN fueling process, the other suppressing it. 
Our data suggest that Sy1-triggering mechanism are generally suppressed 
in CGs, while those triggering Sy2 are activated.  

To investigate whether galaxy-galaxy interaction is responsible for the 
excess of Sy2 in CGs, we have inspected DSS images searching 
for major optical interaction patterns in UZC-CGs. 
We find no Sy1, 3 Sy2 and 7 LINERs that display major interaction patterns.  
The fraction of Sy-CGs with a disturbed Seyfert turns out to be 18\%, 
those with a disturbed LINER nearly 50\%, but the samples are presently 
too small to assess any difference to be significant.  

The fraction of non-Sy-CGs hosting a disturbed/interacting 
galaxy is (51/162) 32\%, and a similar value is found for non-Sy-HCGs. 
Hence, it appears that 
CG presenting obvious interaction patterns 
are not more likely to host a Seyfert galaxy. 
This is consistent with the results of earlier works 
\citep{Moles95,Keel96,Ho97} all finding no clear relationship between 
the presence of nuclear activity and detailed morphological properties. 
  
Recent studies of large samples of Seyfert galaxies have shown 
that Sy and non-Sy are equally likely to present a 
bar \citep{Mulchaey97,Peletier99,Knapen00,Combes03}.   
Among Seyferts in UZC-CGs only 2 (12\%) present a bar. 
The fraction of galaxies with a bar in Sy-CGs and 
non Sy-CGs is 26\% and 18\% respectively, indicating 
that CGs with and without a Seyfert are equally likely to induce 
bars in their member galaxies.
  
Because no evidence is found for Sy-CGs to present higher level of 
galaxy-galaxy interaction, any triggering mechanism responsible for 
the excess of Sy2 observed in CGs does not seem to arise because of the 
strong interaction between two galaxies. 
In general it is difficult, however, to find firm observational evidence 
for dwarf companions because their dynamical disturbance is weak. 
\section{Conclusions}
We have investigated the occurrence of Seyfert galaxies in a nearby sample 
of 192 UZC-CGs. We find 17 Seyferts among the 639 member galaxies, 
indicating that only a minor fraction ($\sim$ 3\%) of UZC-CG galaxies 
host a Seyfert. 

When comparing velocity dispersion, number of large scale neighbours 
and morphological content of Sy-CGs and non-Sy-CGs,  
no significant differences are found, although some Seyferts (5 Sy2) 
appear associated with 'extreme' CGs, presenting large $\sigma$ 
($>$400\,km\,s$^{-1}$), 
many neighbours and an unusually high number of Elliptical members. 
  
This suggests that Sy-CGs and non-Sy-CGs are drawn 
from the same parent population, and that the presence of the Seyfert 
is not linked to specific CG properties. 

We also find the fraction of Sy in CGs (3\%) to be significantly higher 
than the fraction of Sy in a single-galaxy sample (1\%). 
Curiously, the enhanced fraction of Seyferts in CGs reflects the behaviour 
of Sy2 alone: while 14 Sy2 and 3 Sy1 (5:1) are found in UZC-CGs (the ratio is 
6:1 in HCGs), in the single-galaxy sample there are 33 Sy2 and 22 Sy1 (3:2), 
suggesting that the location within a group potential and the triggering of 
type 2 Seyferts are related. 
 
No excess of interacting galaxies or barred galaxies is found among Sy-CGs, 
indicating that strong galaxy-galaxy interaction is not a common Seyfert 
fueling mechanism \citep{Derobertis98,shlosman90} in CGs. 
But any minor merging (between a galaxy and a satellite/dwarf companion) 
fueling mechanism \citep{taniguchi99} is not ruled out. The large number  
of isolated Seyferts we find in our analysis, indirectly supports 
the minor-merging mechanism.
\begin{acknowledgements}
We are pleased to thank R.\,de\,Carvalho, A. Iovino, M. Maia, 
G.G.C.\,Palumbo, \& C.N.A.\,Willmer for stimulating discussions and 
suggestions. We also thank the anonymous referee for comments 
and criticisms that improved the scientific content of the paper. 
This work was supported by MIUR. B.K acknowledges a fellowship of  
Bologna University. 
This research has made use of the NASA/IPAC Extragalactic Database (NED) which is operated by the Jet Propulsion Laboratory, California Institute of Technology, under contract with the National Aeronautics and Space Administration 
\end {acknowledgements}

{}
%%%%%%%%%%%%%%%%%
\end{document}